\documentclass{emulateapj}
\RequirePackage{lineno}
\usepackage{amssymb}
\usepackage{amsmath}
\usepackage{mathrsfs}
\usepackage{natbib}
\usepackage[colorlinks, linkcolor=blue, urlcolor=blue, citecolor=blue]{hyperref}
\usepackage{array}
\usepackage{float}
\usepackage{graphicx}
\usepackage{subfigure}
\usepackage{color}
\usepackage[all]{hypcap}
\usepackage[toc,page]{appendix}
\usepackage{diagbox}
\usepackage{booktabs}
\usepackage{multirow}

\def\beq{\begin{equation}}
\def\enq{\end{equation}}
\def\bea{\begin{eqnarray}}
\def\ena{\end{eqnarray}}

\begin{document}

\title{Determining the efficiency of converting magnetar spin-down energy into gamma-ray burst X-ray afterglow emission and its possible implications}
\author{Di Xiao\altaffilmark{1,2} and Zi-Gao Dai\altaffilmark{1,2}}
\affil{\altaffilmark{1}School of Astronomy and Space Science, Nanjing University, Nanjing 210093, China; dxiao@nju.edu.cn; dzg@nju.edu.cn}
\affil{\altaffilmark{2}Key Laboratory of Modern Astronomy and Astrophysics (Nanjing University), Ministry of Education, China}

\begin{abstract}
Plateaus are common in X-ray afterglows of gamma-ray bursts. Among a few scenarios for the origin of them, the leading one is that there exists a magnetar inside and persistently injects its spin-down energy into an afterglow. In previous studies, the radiation efficiency of this process is assumed to be a constant $\gtrsim0.1$, which is quite simple and strong. In this work we obtain the efficiency from a physical point of view and find that this efficiency strongly depends on the injected luminosity. One implication of this result is that those X-ray afterglow light curves which show steeper temporal decay than $t^{-2}$ after the plateau phase can be naturally understood now. Also, the braking indexes deduced from afterglow fitting are found to be larger than those in previous studies, which are more reasonable for newborn magnetars.
\end{abstract}


\section{Introduction}
After tens-of-years study on gamma-ray bursts (GRBs), one mainstream viewpoint among the community is that there is a dichotomy in their central engines: either black-hole (BH) accretion systems or newborn millisecond magnetars can power GRBs under certain circumstances \citep[for a recent review, see][]{Kumar2015} . Unlike a BH system that usually extracts the gravitational energy of accreted matter \citep[e.g.,][]{Woosley1993, Popham1999, Narayan2001}, a millisecond magnetar extracts its stellar rotational energy to power a GRB and its afterglow \citep[e.g.,][]{Usov1992, Thompson1994, DaiZG1998a, DaiZG1998b, ZhangB2001, Mazzali2014, Beniamini2017c}. Ever since the launch of the {\it Swift} satellite, X-ray afterglows of dozens of GRBs have been found to exhibit plateau features, which are thought to be the signature of a long-lasting energy injection from the central engine \citep{ZhangB2006}. If the central engine is a BH, the injected energy could come from the fall-back accretion onto the BH \citep{Ruffert1997,Rosswog2003, LeiWH2013, WuXF2014}\footnote{Note that fall-back accretion around a magnetar was also proposed to power GRBs \citep[e.g.,][]{Metzger2018}.}. However, the required fall-back mass in the BH model may be too large to explain a plateau at late times ($>10^5$ s) \citep{LiuT2017}. An alternative way is to introduce an spin-down energy injection from a newborn magnetar \citep{DaiZG1998a, DaiZG1998b, ZhangB2001},  which will be discussed in detail below.

The origin of plateaus in X-ray afterglow light curves is still under debate, and basically we would expect two different kinds of them. The first kind is called ``external plateaus" that originate from external shocks. In this case the energy injection comes from a late kinetic-energy-dominated shell interacting with a preceding expanding fireball \citep[e.g.,][]{Rees1998,Panaitescu1998}, so the X-ray light curve should be related to those of other wavelengths \citep[e.g.,][]{Dermer2007, Genet2007, Uhm2007}. The second one is ``internal plateaus" that could reflect the activity of central engines \citep[e.g.,][]{Troja2007, YuYW2009, YuYW2010, Beniamini2017a}. The most prominent feature of an internal plateau is that there is a rapid decay at the end of the plateau, usually with a temporal slope steeper than -3 \citep{LiangEW2007, Lyons2010, Rowlinson2010}. This sudden drop is hard to interpret with a BH central engine but can be well explained as the central magnetar collapse to a BH \citep{Troja2007, Rowlinson2010,Rowlinson2013, LvHJ2014}.

Specifically within the magnetar framework, by what means a magnetar can convert its spin-down energy into radiation is unsure \citep{Usov1999,ZhangB2002}. If the X-ray plateau is ``external", one commonly-discussed physical model is that an accelerated magnetar wind (which is ultra-relativistic, electron-position-pair dominated) interacts with a preceding expanding fireball or an ambient medium \citep{DaiZG2004}. A relativistic ``wind bubble" (which is a relativistic version of pulsar wind nebula) is formed and the reverse shock can accelerate electrons to produce multi-wavelength emission \citep{YuYW2007}. If the X-ray plateau is ``internal",  it can be produced by an internal energy dissipation in the magnetar wind \citep{Coroniti1990, Usov1994}. In this case the spin-down power is mediated by an initially-cold, Poynting-flux-dominated wind that can be gradually accelerated as its magnetic energy dissipates internally via magnetic reconnection \citep{Spruit2001, Drenkhahn2002a, Drenkhahn2002b}. There will be high-energy emission in this process \citep{Giannios2005, Metzger2011, Giannios2012, Beniamini2014, Beniamini2017b, XiaoD2017, XiaoD2018} that can be responsible for the X-ray plateau. In this work we focus on the latter case and calculate the X-ray radiation efficiency in this physical model.

A newborn magnetar loses its rotational energy via gravitational-wave and electromagnetic radiation, whose angular velocity evolution can be generalized as follows \citep{Lasky2017},

\beq
\dot{\Omega}=-k\Omega^n,
\label{eq1}
\enq
where $\Omega=\Omega(t)=2\pi/P(t)$ is the spin angular velocity, and $k$ and $n$ represent a constant of proportionality and the braking index of magnetar respectively. The solution of Eq.(\ref{eq1}) is \citep{Lasky2017, LvHJ2018}
\beq
\Omega(t)=\Omega_0\left(1+\frac{t}{\tau}\right)^{\frac{1}{1-n}}
\label{eq2}
\enq
where $\Omega_0$ is the initial angular velocity and $\tau\equiv\Omega_0^{1-n}/[(n-1)k]$ is the spin-down timescale. The injected energy into the afterglow comes from the magnetic dipole torque whose luminosity is $L_{\rm EM}=B^2R^6\Omega^4/6c^3=L_0(1+t/\tau)^{4/(1-n)}$, where $L_0\equiv B^2R^6\Omega_0^4/6c^3=1.0\times10^{49}B_{15}^{2}R_6^6P_{-3}^{-4}\,\rm erg\,s^{-1}$. Throughout this paper the notation $Q=10^xQ_x$ in cgs units is adopted and the radius of magnetar is assumed to be $R=10^6\,\rm cm$. The observed X-ray plateau luminosity is $L_X=\eta_X L_{\rm EM}$ by introducing an efficiency $\eta_X$, where $\eta_X$ could evolve with time. A bunch of X-ray afterglow light curves with plateau features have been well fitted within the magnetar energy injection scenario, however, all by assuming that $\eta_X$ is constant \citep[e.g.,][]{Lasky2017, LvHJ2018}. We here think better of this assumption in this work.

This paper is organised as follows. In section \ref{sec2} we calculate the X-ray radiation efficiency and obtain its relation on the injected luminosity. Section \ref{sec3} presents the impact of the above relation on afterglow fitting, including both theoretical analysis and case fitting. We also compare our results with previous studies. We finish with conclusions and discussions in Section \ref{sec4}.

\section{X-ray radiation Efficiency}
\label{sec2}
\begin{figure}
\label{fig1}
\begin{center}
\includegraphics[width=0.45\textwidth]{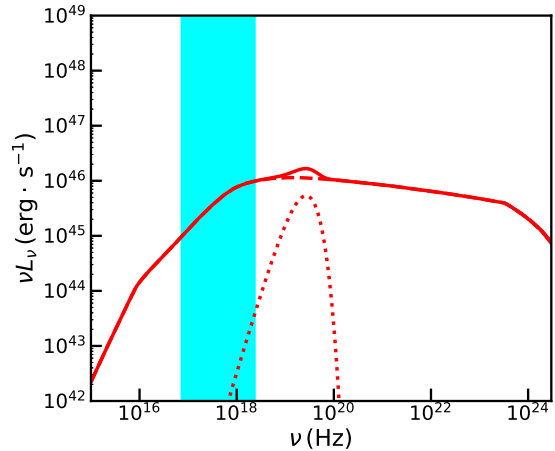}
\caption{The high-energy emission spectrum of the internal gradual magnetic energy dissipation process from the wind of a newborn magnetar with spin period $P=1\,\rm ms$ and magnetic field strength $B=10^{15}\,\rm G$. The total spectrum (red solid line) is a sum of thermal (red dotted line) and non-thermal component (red dashed line). Cyan region represents the observational frequency range of {\it Swift}-XRT.}
\end{center}
\end{figure}

 The wind from a newborn rapidly-rotating magnetar is initially cold and Poynting-flux-dominated \citep{Coroniti1990, Aharonian2012}. As the wind propagates outward, its magnetic energy gradually dissipates via reconnection and is finally converted to high-energy radiation and kinetic energy of the wind. This emission is composed of a thermal component and a non-thermal synchrotron component that can be calculated in detail \citep{Beniamini2017b, XiaoD2017, XiaoD2018}. As an example, Figure \ref{fig1} shows the spectrum of high-energy emission from the newborn magnetar wind with an initial magnetization $\sigma_0=100$. The spin period of central magneter is assumed as $P=1\,\rm ms$ and the magnetic field strength is $B=10^{15}\,\rm G$. Then, the X-ray luminosity can be obtained by integrating on {\it Swift}-XRT band ($0.3-10\,\rm keV$) and the X-ray radiation efficiency is defined as

\beq
\eta_X\equiv\frac{\int_{\rm 0.3\,keV}^{\rm 10\,keV}L_\nu d\nu}{L_{\rm EM}}\label{eq3}.
\enq

\begin{figure}
\label{fig2}
\begin{center}
\includegraphics[width=0.45\textwidth]{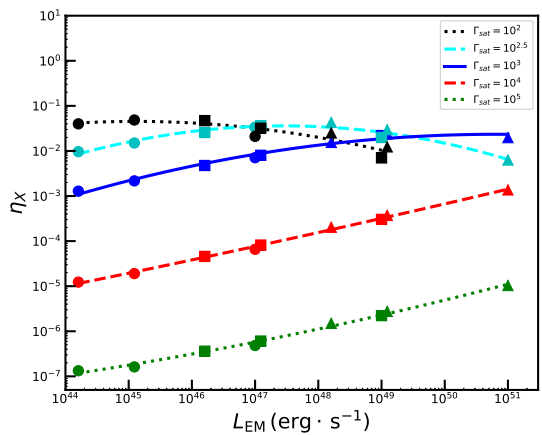}
\caption{Polynomial fitting of the dependences of $\eta_X$ on the injected luminosity $L_{\rm EM}$. We consider five cases of $\Gamma_{\rm sat}=10^2,\,10^{2.5},\,10^3,\,10^4,\,10^5$ that are indicated in the upper-right corner. Different symbols are used to differentiate the magnetic field strength: circles, squares and triangles are for $B=10^{14},\,10^{15}$ and $10^{16}\,\rm G$ respectively.}
\end{center}
\end{figure}

Further, we can calculate efficiencies with other given values of $P$ and $B$. For different parameter sets, different cases are named in the form of ``P$x$B$y$'', with $x$ denoting the spin period in ms and $y$ denoting the logarithm of the magnetic field strength in Gauss. Other than the spin and magnetic field, the wind's saturation Lorentz factor $\Gamma_{\rm sat}$ that is related to the initial magnetization parameter $\sigma_0$ as $\Gamma_{\rm sat}=\sigma_0^{3/2}$ \citep{Beniamini2017b} also plays an important role, which has been discussed in \citet{XiaoD2017}. We have calculated the efficiencies for parameter assemblies within $P=1,\,3,\,5\,\rm ms$, $B=10^{14},\,10^{15},\,10^{16}\,\rm G$ and $\Gamma_{\rm sat}=10^2,\,10^{2.5},\,10^3,\,10^4,\,10^5$. The results are listed in Table \ref{table1}. We carry out polynomial fitting to obtain the dependences of $\eta_X$ on $L_{\rm EM}$, which are
\bea
\log\eta_X = -0.042(\log L_{\rm EM})^2 + 3.81\log L_{\rm EM}-87.29, \nonumber\\
\log\eta_X = -0.059(\log L_{\rm EM})^2 + 5.61\log L_{\rm EM}-134.60, \nonumber\\
\log\eta_X = -0.030(\log L_{\rm EM})^2 + 3.10\log L_{\rm EM}-80.28, \nonumber\\
\log\eta_X = -0.0034(\log L_{\rm EM})^2 - 0.016\log L_{\rm EM}-10.89, \nonumber\\
\log\eta_X = -0.011(\log L_{\rm EM})^2 - 0.75\log L_{\rm EM}+4.78, \nonumber\\
\label{eta}
\ena
for $\Gamma_{\rm sat}=10^2,\,10^{2.5},\,10^3,\,10^4,\,10^5$ respectively, as shown in Figure \ref{fig2}. As we can see clearly, the X-ray efficiency strongly depends on the injected luminosity $\eta_X=\eta_X(L_{\rm EM})$, which will influence the X-ray light curve at late times.

\renewcommand\arraystretch{2}
\begin{table*}
\centering
\caption{X-ray radiation efficiencies using different parameter sets}
\vspace{2mm}
\begin{tabular}{c|ccccccccc}
\hline
\hline
\diagbox[dir=SE]{$\Gamma_{\rm sat}$}{$\eta_X$}{P, B} & P5B14 & P3B14 & P1B14 & P5B15 & P3B15 & P1B15 & P5B16 & P3B16 & P1B16 \\
\hline
$\Gamma_{\rm sat}=10^2$ & $3.99\times10^{-2}$ & $4.86\times10^{-2}$ & $2.10\times10^{-2}$ & $4.72\times10^{-2}$ & $3.20\times10^{-2}$ & $7.00\times10^{-3}$ & $2.51\times10^{-2}$ & $1.23\times10^{-2}$ & $-\footnote{For this parameter set the saturation radius is even smaller than the photospheric radius. The high-energy emission is then totally thermalized and the model in this work does not apply.}$ \\
\hline
$\Gamma_{\rm sat}=10^{2.5}$ & $9.63\times10^{-3}$ & $1.50\times10^{-2}$ & $3.36\times10^{-2}$ & $2.57\times10^{-2}$ & $3.65\times10^{-2}$ & $1.99\times10^{-2}$ & $4.32\times10^{-2}$ & $2.96\times10^{-2}$ & $6.30\times10^{-3}$ \\
\hline
$\Gamma_{\rm sat}=10^3$ & $1.28\times10^{-3}$ & $2.17\times10^{-3}$ & $7.08\times10^{-3}$ & $4.76\times10^{-3}$ & $8.01\times10^{-3}$ & $2.20\times10^{-2}$ & $1.54\times10^{-2}$ & $2.40\times10^{-2}$ & $1.98\times10^{-2}$ \\
\hline
$\Gamma_{\rm sat}=10^4$ & $1.23\times10^{-5}$ & $1.90\times10^{-5}$ & $6.58\times10^{-5}$ & $4.57\times10^{-5}$ & $8.18\times10^{-5}$ & $3.03\times10^{-4}$ & $2.03\times10^{-4}$ & $3.72\times10^{-4}$ & $1.37\times10^{-3}$ \\
\hline
$\Gamma_{\rm sat}=10^5$ & $1.33\times10^{-7}$ & $1.62\times10^{-7}$ & $4.83\times10^{-7}$ & $3.62\times10^{-7}$ & $6.07\times10^{-7}$ & $2.22\times10^{-6}$ & $1.50\times10^{-6}$ & $2.76\times10^{-6}$ & $1.03\times10^{-5}$ \\
\hline
\hline
\end{tabular}
\label{table1}
\end{table*}

\section{Impact on the temporal decay index after plateau}
\label{sec3}
\subsection{Theoretical Analysis}
\begin{figure}
\label{fig3}
\begin{center}
\includegraphics[width=0.45\textwidth]{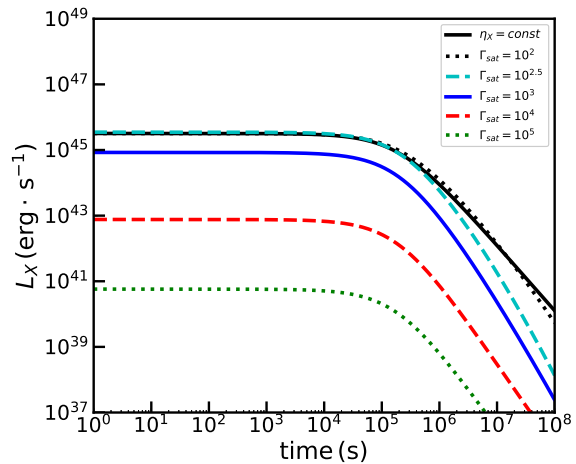}
\caption{The theoretical X-ray light curves of magnetar energy injection. Different lines represent different $\Gamma_{\rm sat}$ indicated in the upper-right corner. The black solid line represent the traditional $\eta_X=\rm constant$ assumption and it has been normalized to the initial value of $\Gamma_{\rm sat}=10^2$ case. Clearly the temporal decay indexes after plateau of all five cases deviate from $\beta=-2$.}
\end{center}
\end{figure}

Since the efficiency evolves as the injected electromagnetic luminosity decreases, we can expect that the decay index $\beta$ of X-ray flux after the plateau phase ($F_X\propto t^{\beta}$) will deviate from the commonly-believed value $-2$. \citep[e.g.,][]{ZhangB2001}. Figure \ref{fig3} shows how the X-ray light curves behave if we take the relation $\eta_X=\eta_X(L_{\rm EM})$ into account. For three cases of $\Gamma_{\rm sat}=10^3,\,10^4,\,10^5$, since $\eta_X$ decreases monotonously with decreasing $L_{\rm EM}$, the temporal indexes appear $\beta<-2$ after plateau. However, for $\Gamma_{\rm sat}=10^2,\,10^{2.5}$ cases, $\eta_X$ first increases and then decreases later with decreasing $L_{\rm EM}$. This leads to $\beta>-2$ above a critical value $L_{\rm EM, cr}$ and turns into $\beta<-2$ after $L_{\rm EM}$ drops below this value. This break of temporal decay index is totally caused by evolution of $\eta_X$ and an application of this effect to individual cases needs fine tuning and are left for future work.

In the conventional picture, if the magnetar spins down only through a dipole torque, the braking index $n=3$. If it spins down only through gravitational-wave radiation, then $n=5$ \citep{Shapiro1983}. For a newborn magnetar, we can expect these two mechanisms are both very important and their combined effect on spin evolution leads to $3\leq n\leq5$ \footnote{A larger range of the braking index is possible if the other mechanism is taken into account. For instance, neutron stars that spin down through unstable r-modes have $n=7$ \citep{Owen1998}. Also, the fall-back accretion could lead to $n<3$ \citep{Metzger2018}.}. As long as $k$ is constant in Eq.(\ref{eq1}), we can deduce from Eq.(\ref{eq2}) that the decay index of X-ray light curve after plateau is $\beta=4/(1-n)$ and should lie between -1 and -2. However, observationally we have found a lot of cases with decay indexes $\beta<-2$. Traditionally we have to make a further assumption that $k$ evolves with time to reconcile this discrepancy \citep{Lasky2017}. However, we have shown here that the evolution of $\eta_X$ with time is a more natural explanation that should be given priority to.

\begin{figure}
\centering
\subfigure[The fitting result of X-ray afterglow light curve for GRB 100615A.]{
\begin{minipage}[b]{0.45\textwidth}
\includegraphics[width=1\textwidth]{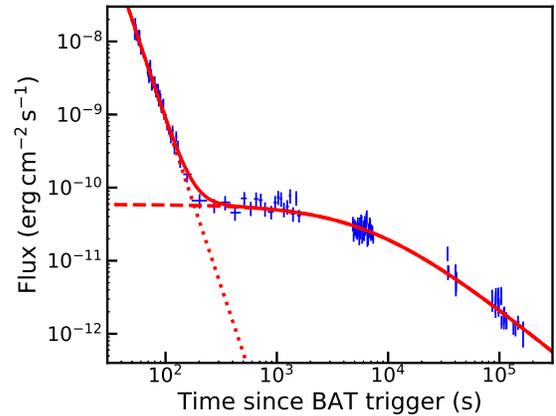}
\label{fig4:subfig:a}
\end{minipage}
}
\subfigure[The fitting result of X-ray afterglow light curve for GRB 150910A.]{
\begin{minipage}[b]{0.45\textwidth}
\includegraphics[width=1\textwidth]{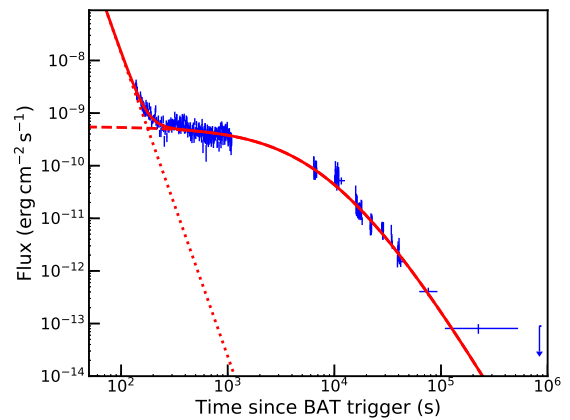}
\label{fig4:subfig:b}
\end{minipage}
}
\subfigure[The X-ray efficiencies as a function of time for the above two cases: blue for GRB 100615A and green for GRB 150910A.]{
\begin{minipage}[b]{0.45\textwidth}
\includegraphics[width=1\textwidth]{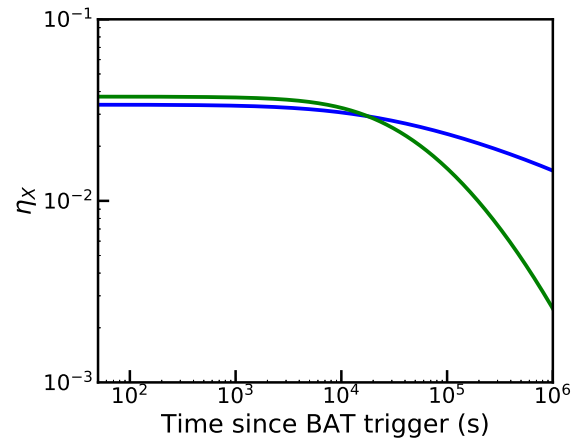}
\label{fig4:subfig:c}
\end{minipage}
}
\caption{Two examples of afterglow fitting results: GRB 100615A with $\beta > -2$ and GRB 150910A with $\beta<-2$.}
\label{fig4}
\end{figure}

\begin{figure*}
\label{fig5} \centering\includegraphics[angle=0,height=7.5in]{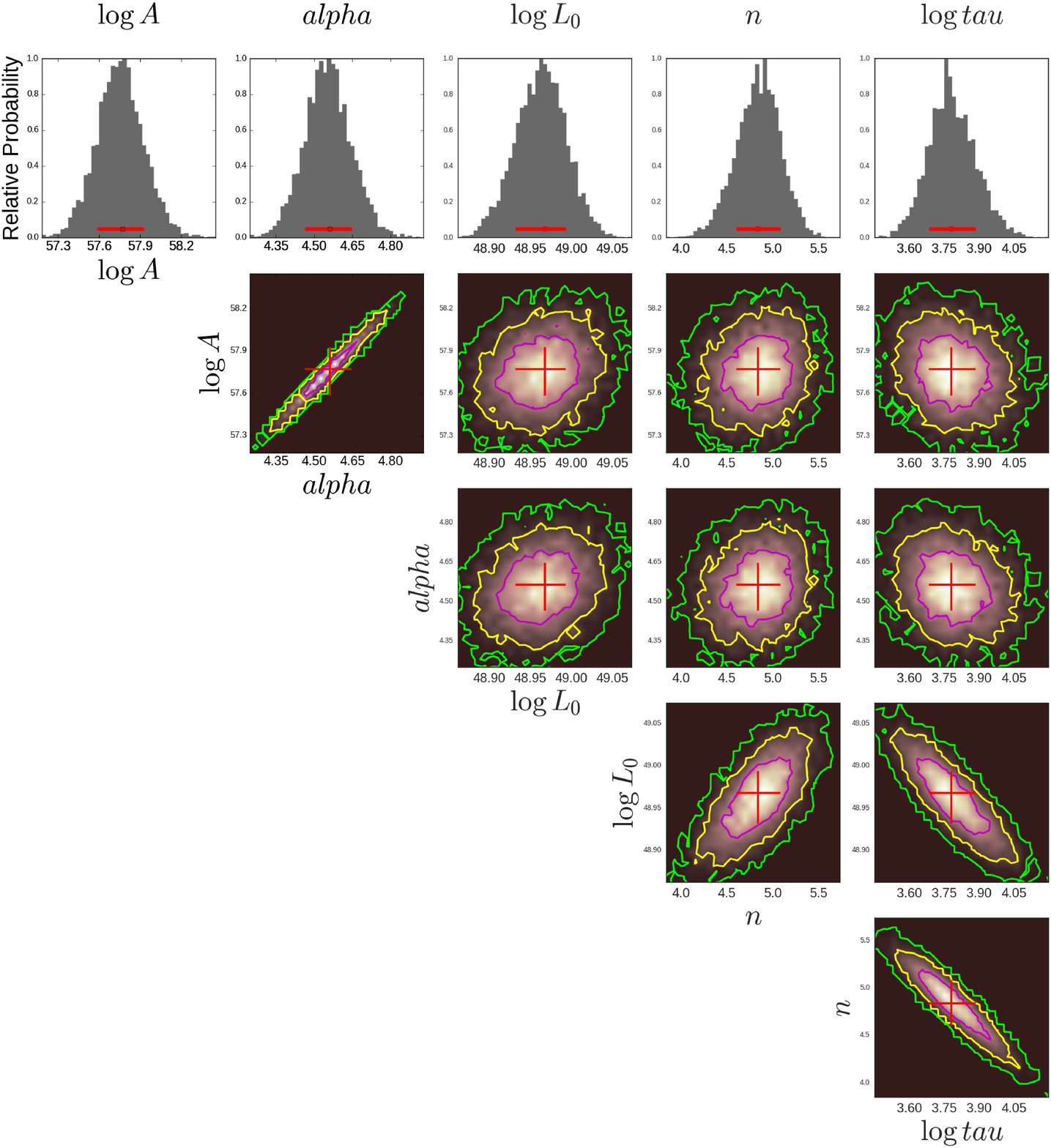} \ \
\caption{Parameter constraints of afterglow light curve fitting for GRB 100615A. Histograms and contours illustrate the likelihood map. Red crosses show the best-fitting values and
1-sigma error bars.}
\end{figure*}

\begin{figure*}
\label{fig6} \centering\includegraphics[angle=0,height=7.5in]{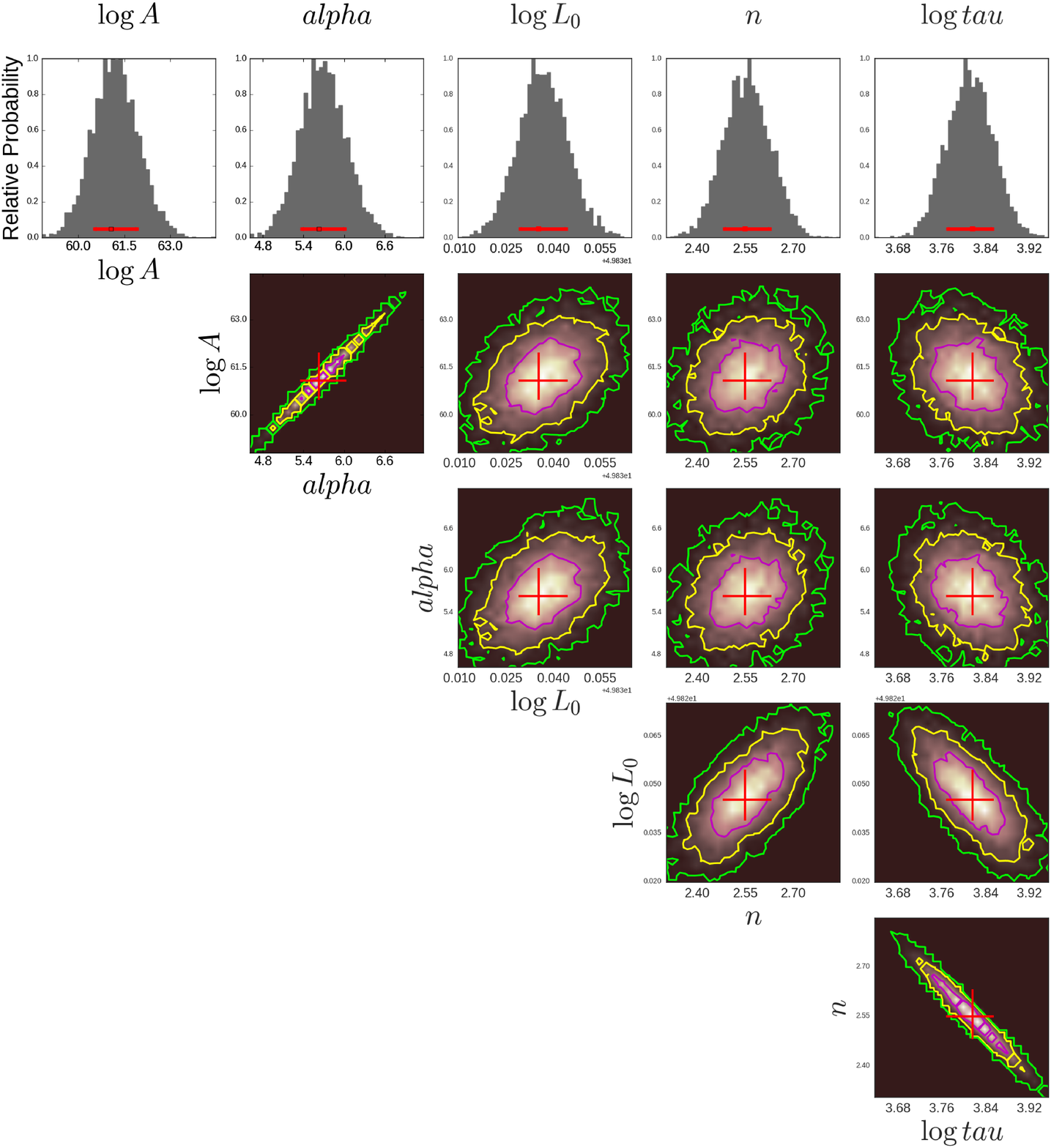} \ \
\caption{The same as Figure 5 but for GRB 150910A.}
\end{figure*}

We can now apply our result to fit individual cases. The initial steep decay of X-ray afterglow light curve is fitted with a power-law component $L_{\rm pl}=At^{-\alpha}$ and the observed X-ray flux is then $F_X=(1+z)(\eta_XL_{\rm EM}+L_{\rm pl})/4\pi D_L^2$, where $z$ is redshift and $D_L$ is the corresponding luminosity distance. Note that at different redshifts the ranges of integration of Eq.(\ref{eq3}) in the burst frame vary, so that $\eta_X$ should be calculated case by case. Taking $(A,\,\alpha,\,L_0,\,n,\,\tau)$ as parameters we can do a Bayesian Monte-Carlo fitting using MCurveFit package \citep{ZhangBB2016}. Figure \ref{fig4} gives two examples of afterglow fitting results, that are GRB 100615A with normal $\beta > -2$ and GRB 150910A with $\beta<-2$. We can see from Figure \ref{fig4:subfig:c} that the X-ray efficiencies are smaller than 0.1 and trace the time evolution of $L_{\rm EM}$. For this figure and the results below, we assume $\Gamma_{\rm sat}=10^3$, which corresponds to initial magnetization $\sigma_0=100$ and is very typical for a GRB \citep{Beniamini2017b}. The best fitting values and parameter corners are shown in Table \ref{table2} and Figure \ref{fig5} \& \ref{fig6}. Since the power law component can be well identified, once we fix $A$ or $\alpha$ there is not much space for the other, so $A$ and $\alpha$ is highly degenerated. Also, as we can see from Eq.(\ref{eq2}) and the definition below it, there is a correlation among $L_0$, $\tau$ and $n$. Therefore, either two of them could be moderately degenerated.

\begin{table}
\centering
\caption{The best-fitting values for the five parameters using Bayesian Monte-Carlo method.}
\label{table2}
\subtable[Parameters for GRB 100615A]{
\begin{tabular}{ccc}
\toprule Parameter &Allowed range & Best-fitting value \\
\midrule$\log A$ &[$40,\,100$]& $57.77_{-0.19}^{+0.16}$ \\
$\alpha$ &[0, 15] & $4.56_{-0.098}^{+0.084}$ \\
$\log L_0$ &[42, 52] & $48.97_{-0.036}^{+0.025}$ \\
$n$ &[1, 7]& $4.84_{-0.24}^{+0.24}$ \\
$\log\tau$ &[$1,\,10$]& $3.78_{-0.098}^{+0.11}$ \\
\bottomrule & &
\end{tabular}
}
\subtable[Parameters for GRB 150910A]{
\begin{tabular}{ccc}
\toprule Parameter &Allowed range & Best-fitting value \\
\midrule$\log A$ &[$40,\,100$]& $61.07_{-0.60}^{+0.88}$ \\
$\alpha$ &[0, 15] & $5.63_{-0.27}^{+0.40}$ \\
$\log L_0$ &[42, 52] & $49.87_{-0.0065}^{+0.0093}$ \\
$n$ &[1, 7]& $2.55_{-0.070}^{+0.082}$ \\
$\log\tau$ &[$1,\,10$]& $3.82_{-0.048}^{+0.038}$ \\
\bottomrule & &
\end{tabular}
}
\end{table}

\subsection{Comparison with Previous Work}
\begin{figure*}
\label{fig7}
\begin{center}
\includegraphics[angle=0,height=4.5in]{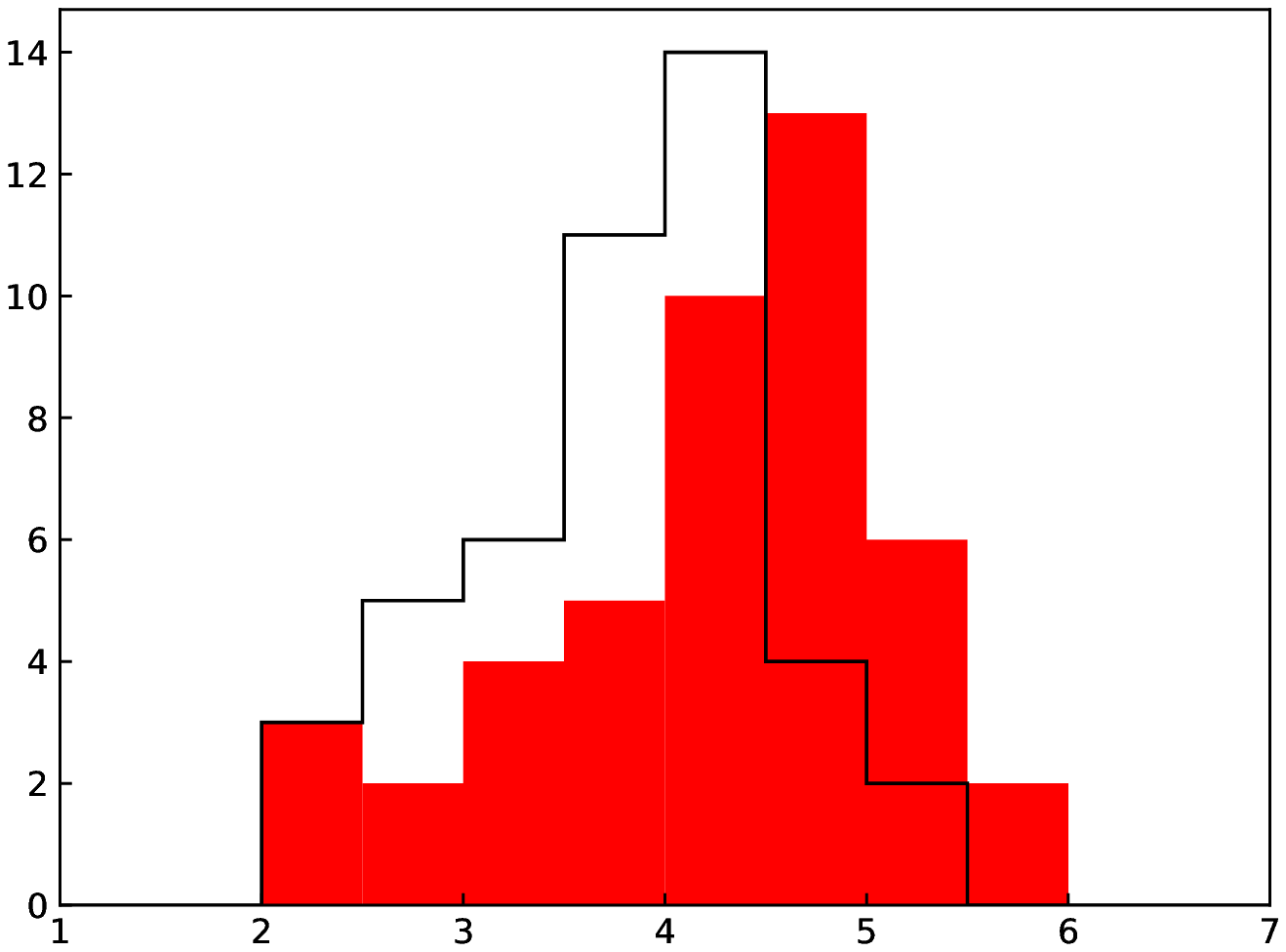}
\caption{The comparison on the distribution of braking index between \citet{LvHJ2018} (black hollow histogram) and this work (red filled histogram).}
\end{center}
\end{figure*}

Since the X-ray light curve decays faster in our analysis, the braking index deduced from afterglow fitting should also differ from previous studies. For example if $\beta=-2$, traditionally we get $n=3$. However, as long as $\eta_X$ decreases with time, in our analysis $4/(1-n)>-2$ is required, leading to $n>3$. Therefore, generally we will obtain larger values of $n$ from afterglow fitting. In order to illustrate this effect directly, we adopt the same sample as in \citet{LvHJ2018} and compare our result with theirs, in which constant $\eta_X=0.1$ was assumed. GRB 100615 was within that sample and their best-fitting braking index is $n=4.61\pm0.14$, which is smaller than our value in Table \ref{table2}. This is easy to understand from the above discussion. In fact, we have redone the fitting using our relation and a complete comparison of fitted parameters with \citet{LvHJ2018} is shown in Table \ref{table3} and the distribution of $n$ is shown in Figure \ref{fig7}. Since our fitted X-ray efficiency is less than 0.1, the values of $L_0$ are universally larger than those in \citet{LvHJ2018}. Moreover, we find that the deduced braking indexes are also universally larger in this work. The central position of the distribution on $n$ is also shifted to a larger value.

A critical point we should check is whether the values of parameters chosen for fitting the X-ray plateaus are consistent with the luminosity requirements during the prompt phase of the same bursts. As we can see in Table 3, the deduced initial injected luminosity $L_0$ is centered near $10^{49} \,\rm erg\,s^{-1}$, which looks smaller than the typical prompt luminosity. However, an important uncertainty should be taken into account, which is the beaming effect. At such late times of X-ray plateau, the outflow has gone through remarkable sideways expansion and the injected luminosity is quasi-isotropic. But during the early prompt phase, the jet should be highly beamed. If we assume a double-sided jet with an opening angle $\sim 0.1$, the observed luminosity during prompt phase is then 200 times higher, which is of order $10^{51} \,\rm erg\,s^{-1}$ and can easily match the luminosity requirement of the prompt phase. Furthermore, there will be some correlation for the observed spectrum between the prompt and X-ray plateau phases. As we have discussed in \citet{XiaoD2018}, the temperature of the thermal component depends weakly on the injected luminosity ($T_{\rm ph} \propto L_{\rm EM}^{1/10}$ from Eq.(3) in that work). Therefore, even the luminosity is 200 times lower during X-ray plateau phase due to jet widening, the temperature is only 1.7 times lower than that of the prompt phase. Thus, we can expect that the peak energy is nearly the same in these two phases. Note that the above discussion is valid only if the prompt and X-ray plateau emission both originate from the gradual magnetic dissipation process. If they do not have the same origin mechanism, there will be no correlation of the peak energy during these two phases. For example, if the prompt emission comes from the accretion of a newborn magnetar \citep[e.g.,][]{ZhangD2008,ZhangD2009,ZhangD2010} or the differential rotation in the magnetar's interior \citep[e.g.,][]{Kulzniak1998,DaiZG1998b}, then its luminosity and spectral properties depend on the mass accretion rate or the differentially rotational energy.

\renewcommand\arraystretch{2}
\begin{table*}
\centering
\caption{The comparison on the parameters deduced in this work and \citet{LvHJ2018}.}
\resizebox{0.9\textwidth}{120mm}{
\begin{tabular}{ccccccccc}
\hline
\hline
\multirow{2}{*}{GRB} & \multirow{2}{*}{$\log A$} & \multirow{2}{*}{$\alpha$} & \multicolumn{2}{c}{$\log L_0$} & \multicolumn{2}{c}{n}  & \multicolumn{2}{c}{$\log\tau$} \\
\cline{4-5}\cline{6-7}\cline{8-9} & & & this work & \citet{LvHJ2018} & this work & \citet{LvHJ2018}& this work & \citet{LvHJ2018}\\\hline
050319 & $53.79_{-0.77}^{+6.63}$ & $2.20_{-0.30}^{+2.62}$ & $48.92_{-0.04}^{+0.11}$ & $47.81\pm0.01$ & $3.88_{-0.25}^{+0.50}$ & $4.14\pm0.14$ & $4.25_{-0.28}^{+0.10}$ & $3.87\pm0.04$\\
050822 & $74.97_{-2.56}^{+3.22}$ & $9.72_{-0.92}^{+1.17}$ & $48.30_{-0.05}^{+0.03}$ & $46.65\pm0.01$ & $5.35_{-0.15}^{+0.12}$ & $4.45\pm0.11$ & $3.92_{-0.06}^{+0.10}$ & $4.12\pm0.04$\\
050922B & $77.54_{-0.77}^{+1.06}$ & $9.44_{-0.25}^{+0.35}$ & $48.12_{-0.04}^{+0.04}$ & $46.87\pm0.02$ & $3.30_{-0.28}^{+0.36}$ & $2.94\pm0.26$ & $5.29_{-0.15}^{+0.11}$ & $5.30\pm0.14$\\
051016B & $61.28_{-1.66}^{+1.70}$ & $6.89_{-0.84}^{+0.86}$ & $47.87_{-0.04}^{+0.04}$ & $46.12\pm0.001$ & $4.67_{-0.19}^{+0.16}$ & $3.80\pm0.13$ & $3.98_{-0.08}^{+0.09}$ & $4.14\pm0.05$\\
060604 & $70.71_{-0.58}^{+0.90}$ & $9.39_{-0.25}^{+0.39}$ & $48.37_{-0.04}^{+0.05}$ & $47.02\pm0.02$ & $4.51_{-0.15}^{+0.22}$ & $4.04\pm0.13$ & $4.06_{-0.12}^{+0.08}$ & $4.03\pm0.05$\\
060605 & $53.74_{-0.82}^{+1.78}$ & $2.58_{-0.39}^{+0.86}$ & $49.18_{-0.05}^{+0.05}$ & $48.04\pm0.01$ & $2.47_{-0.17}^{+0.15}$ & $2.48\pm0.08$ & $4.09_{-0.10}^{+0.11}$ & $3.95\pm0.04$\\
060714 & $81.45_{-1.38}^{+0.85}$ & $14.05_{-0.61}^{+0.37}$ & $49.16_{-0.04}^{+0.02}$ & $47.78\pm0.02$ & $4.46_{-0.16}^{+0.08}$ & $3.86\pm0.10$ & $3.34_{-0.05}^{+0.10}$ & $3.49\pm0.05$\\
060729 & $60.94_{-0.09}^{+0.09}$ & $5.39_{-0.04}^{+0.04}$ & $47.99_{-0.01}^{+0.01}$ & $46.24\pm0.0005$ & $3.94_{-0.04}^{+0.05}$ & $3.34\pm0.03$ & $4.92_{-0.02}^{+0.01}$ & $4.96\pm0.01$\\
061121 & $60.71_{-0.10}^{+0.13}$ & $5.52_{-0.05}^{+0.06}$ & $49.54_{-0.01}^{+0.01}$ & $48.16\pm0.01$ & $4.09_{-0.03}^{+0.04}$ & $3.71\pm0.03$ & $3.47_{-0.02}^{+0.02}$ & $3.45\pm0.01$\\
070129 & $70.22_{-0.25}^{+0.31}$ & $7.57_{-0.09}^{+0.11}$ & $48.22_{-0.03}^{+0.02}$ & $46.85\pm0.01$ & $4.40_{-0.17}^{+0.13}$ & $3.95\pm0.12$ & $4.40_{-0.06}^{+0.08}$ & $4.31\pm0.04$\\
070306 & $64.64_{-0.29}^{+0.21}$ & $6.67_{-0.12}^{+0.09}$ & $48.66_{-0.01}^{+0.01}$ & $47.17\pm0.01$ & $2.69_{-0.12}^{+0.09}$ & $2.38\pm0.08$ & $4.81_{-0.04}^{+0.06}$ & $4.86\pm0.04$\\
070328 & $63.79_{-1.76}^{+1.80}$ & $7.37_{-0.86}^{+0.91}$ & $50.69_{-0.01}^{+0.01}$ & $49.34\pm0.004$ & $3.51_{-0.03}^{+0.03}$ & $3.33\pm0.02$ & $2.79_{-0.02}^{+0.02}$ & $2.83\pm0.01$\\
070508 & $70.35_{-9.47}^{+5.10}$ & $13.22_{-8.52}^{+0.29}$ & $50.06_{-0.01}^{+0.01}$ & $48.51\pm0.003$ & $4.02_{-0.03}^{+0.03}$ & $3.56\pm0.02$ & $2.64_{-0.02}^{+0.02}$ & $2.79\pm0.01$\\
080430 & $52.99_{-0.30}^{+0.19}$ & $2.62_{-0.15}^{+0.10}$ & $48.00_{-0.03}^{+0.03}$ & $46.44\pm0.02$ & $5.27_{-0.09}^{+0.14}$ & $4.68\pm0.07$ & $3.96_{-0.07}^{+0.06}$ & $3.76\pm0.03$\\
081210 & $54.52_{-0.53}^{+0.87}$ & $2.35_{-0.19}^{+0.32}$ & $48.00_{-0.08}^{+0.14}$ & $46.45\pm0.02$ & $3.93_{-0.65}^{+0.92}$ & $3.34\pm0.33$ & $4.63_{-0.44}^{+0.27}$ & $4.72\pm0.14$\\
090404 & $64.85_{-0.10}^{+0.10}$ & $7.01_{-0.05}^{+0.05}$ & $48.64_{-0.03}^{+0.01}$ & $47.31\pm0.01$ & $4.13_{-0.15}^{+0.08}$ & $3.63\pm0.11$ & $4.25_{-0.03}^{+0.07}$ & $4.30\pm0.04$\\
090516 & $80.96_{-0.98}^{+1.35}$ & $12.44_{-0.39}^{+0.55}$ & $50.85_{-0.10}^{+0.30}$ & $48.05\pm0.01$ & $4.28_{-0.04}^{+0.07}$ & $2.81\pm0.07$ & $2.22_{-0.30}^{+0.10}$ & $4.09\pm0.03$\\
090529 & $57.43_{-0.22}^{+0.18}$ & $3.55_{-0.09}^{+0.08}$ & $47.61_{-0.08}^{+0.11}$ & $46.02\pm0.04$ & $5.25_{-0.74}^{+1.10}$ & $4.03\pm0.62$ & $4.42_{-0.43}^{+0.29}$ & $4.69\pm0.27$\\
090618 & $62.75_{-0.11}^{+0.14}$ & $5.97_{-0.05}^{+0.06}$ & $49.56_{-0.01}^{+0.01}$ & $47.96\pm0.002$ & $4.38_{-0.02}^{+0.02}$ & $3.82\pm0.01$ & $3.04_{-0.02}^{+0.02}$ & $3.16\pm0.01$\\
091018 & $61.86_{-17.69}^{+2.89}$ & $8.17_{-3.23}^{+5.21}$ & $49.65_{-0.01}^{+0.04}$ & $48.13\pm0.01$ & $4.68_{-0.03}^{+0.08}$ & $4.12\pm0.04$ & $2.39_{-0.08}^{+0.02}$ & $2.51\pm0.03$\\
091029 & $63.47_{-1.31}^{+3.56}$ & $6.08_{-0.51}^{+1.41}$ & $48.68_{-0.01}^{+0.03}$ & $47.32\pm0.01$ & $4.65_{-0.08}^{+0.14}$ & $4.08\pm0.09$ & $3.99_{-0.07}^{+0.03}$ & $4.05\pm0.03$\\
100302A & $62.56_{-0.26}^{+0.45}$ & $5.16_{-0.10}^{+0.17}$ & $48.10_{-0.06}^{+0.10}$ & $46.83\pm0.02$ & $5.83_{-0.38}^{+0.48}$ & $5.05\pm0.26$ & $4.03_{-0.25}^{+0.16}$ & $4.06\pm0.09$\\
100615A & $57.77_{-0.19}^{+0.16}$ & $4.56_{-0.10}^{+0.08}$ & $48.97_{-0.04}^{+0.03}$ & $47.55\pm0.01$ & $4.84_{-0.24}^{+0.24}$ & $4.61\pm0.14$ & $3.78_{-0.10}^{+0.11}$ & $3.65\pm0.05$\\
100814A & $60.83_{-0.13}^{+0.21}$ & $4.83_{-0.05}^{+0.09}$ & $48.54_{-0.02}^{+0.02}$ & $47.11\pm0.01$ & $3.24_{-0.10}^{+0.07}$ & $3.02\pm0.06$ & $4.96_{-0.04}^{+0.05}$ & $4.85\pm0.02$\\
110808A & $56.62_{-0.26}^{+0.33}$ & $3.91_{-0.12}^{+0.16}$ & $47.47_{-0.07}^{+0.12}$ & $45.87\pm0.03$ & $5.46_{-0.62}^{+0.79}$ & $4.87\pm0.41$ & $4.30_{-0.37}^{+0.21}$ & $4.13\pm0.15$\\
111008A & $58.42_{-0.14}^{+0.20}$ & $4.13_{-0.07}^{+0.10}$ & $49.61_{-0.02}^{+0.02}$ & $48.41\pm0.01$ & $4.04_{-0.06}^{+0.10}$ & $3.69\pm0.07$ & $3.72_{-0.06}^{+0.03}$ & $3.81\pm0.03$\\
111228A & $60.87_{-0.11}^{+0.10}$ & $5.30_{-0.05}^{+0.05}$ & $48.48_{-0.03}^{+0.02}$ & $46.88\pm0.01$ & $4.57_{-0.08}^{+0.09}$ & $3.97\pm0.06$ & $3.84_{-0.05}^{+0.05}$ & $3.83\pm0.02$\\
120422A & $60.87_{-0.28}^{+0.24}$ & $6.54_{-0.14}^{+0.12}$ & $46.06_{-0.06}^{+0.03}$ & $43.81\pm0.03$ & $4.75_{-1.07}^{+1.02}$ & $3.63\pm0.66$ & $5.13_{-0.26}^{+0.31}$ & $5.22\pm0.27$\\
120521C & $55.82_{-0.18}^{+0.22}$ & $3.24_{-0.09}^{+0.12}$ & $48.31_{-0.05}^{+0.08}$ & $47.27\pm0.03$ & $2.33_{-0.51}^{+1.66}$ & $3.12\pm0.64$ & $4.65_{-0.62}^{+0.29}$ & $4.07\pm0.32$\\
130609B & $66.38_{-0.23}^{+0.34}$ & $6.80_{-0.09}^{+0.13}$ & $49.88_{-0.01}^{+0.01}$ & $48.42\pm0.01$ & $3.02_{-0.05}^{+0.07}$ & $2.72\pm0.04$ & $3.52_{-0.04}^{+0.03}$ & $3.61\pm0.02$\\
131105A & $78.96_{-3.74}^{+3.65}$ & $12.51_{-1.52}^{+1.47}$ & $48.82_{-0.03}^{+0.03}$ & $47.35\pm0.01$ & $4.65_{-0.29}^{+0.19}$ & $4.02\pm0.17$ & $3.51_{-0.08}^{+0.12}$ & $3.60\pm0.07$\\
140430A & $76.29_{-0.40}^{+0.75}$ & $11.32_{-0.17}^{+0.31}$ & $48.29_{-0.12}^{+0.03}$ & $46.75\pm0.02$ &  $5.25_{-0.29}^{+0.05}$ & $4.87\pm0.41$ & $3.77_{-0.02}^{+0.27}$ & $3.75\pm0.12$\\
140703A & $63.62_{-0.18}^{+0.36}$ & $6.36_{-0.09}^{+0.17}$ & $49.51_{-0.04}^{+0.03}$ & $48.32\pm0.01$ & $2.13_{-0.13}^{+0.14}$ & $2.17\pm0.08$ & $4.43_{-0.09}^{+0.09}$ & $4.30\pm0.05$\\
160227A & $58.42_{-0.13}^{+0.15}$ & $3.44_{-0.05}^{+0.06}$ & $48.78_{-0.03}^{+0.03}$ & $47.48\pm0.01$ & $4.15_{-0.17}^{+0.07}$ & $3.77\pm0.10$ & $4.45_{-0.04}^{+0.08}$ & $4.37\pm0.04$\\
160804A & $68.23_{-0.31}^{+0.34}$ & $7.25_{-0.11}^{+0.12}$ & $47.71_{-0.03}^{+0.05}$ & $46.07\pm0.001$ & $5.87_{-0.43}^{+0.40}$ & $5.26\pm0.26$ & $4.00_{-0.14}^{+0.12}$ & $3.80\pm0.07$\\
161117A & $60.83_{-0.08}^{+0.11}$ & $4.96_{-0.04}^{+0.05}$ & $49.07_{-0.04}^{+0.03}$ & $47.58\pm0.01$ & $4.80_{-0.11}^{+0.09}$ & $4.18\pm0.08$ & $3.51_{-0.05}^{+0.06}$ & $3.62\pm0.03$\\
170113A & $64.91_{-0.38}^{+0.37}$ & $7.56_{-0.18}^{+0.18}$ & $49.62_{-0.02}^{+0.03}$ & $48.21\pm0.01$ & $4.71_{-0.09}^{+0.07}$ & $4.17\pm0.07$ & $3.06_{-0.05}^{+0.05}$ & $3.19\pm0.03$\\
170519A & $67.24_{-0.18}^{+0.14}$ & $7.64_{-0.07}^{+0.06}$ & $48.33_{-0.04}^{+0.02}$ & $46.72\pm0.01$ & $3.42_{-0.19}^{+0.11}$ & $2.99\pm0.12$ & $4.06_{-0.05}^{+0.08}$ & $4.07\pm0.05$\\
170531B & $88.03_{-0.92}^{+0.79}$ & $13.97_{-0.33}^{+0.28}$ & $48.45_{-0.06}^{+0.17}$ & $46.98\pm0.03$ & $4.89_{-0.59}^{+1.05}$ & $4.01\pm0.49$ & $3.52_{-0.50}^{+0.19}$ & $3.69\pm0.17$\\
170607A & $54.63_{-0.05}^{+0.06}$ & $2.80_{-0.02}^{+0.02}$ & $48.01_{-0.03}^{+0.02}$ & $46.27\pm0.001$ & $5.16_{-0.11}^{+0.16}$ & $4.43\pm0.11$ & $4.19_{-0.05}^{+0.06}$ & $4.21\pm0.04$\\
170705A & $59.03_{-0.08}^{+0.14}$ & $3.99_{-0.03}^{+0.06}$ & $49.13_{-0.03}^{+0.03}$ & $47.79\pm0.01$ & $4.70_{-0.10}^{+0.09}$ & $4.28\pm0.08$ & $4.01_{-0.06}^{+0.05}$ & $3.98\pm0.03$\\
171222A & $59.09_{-0.06}^{+0.06}$ & $3.92_{-0.02}^{+0.02}$ & $47.52_{-0.06}^{+0.05}$ & $46.06\pm0.04$ & $4.36_{-1.32}^{+0.96}$ & $4.01\pm0.91$ & $5.25_{-0.24}^{+0.39}$ & $5.10\pm0.36$\\
180325A & $63.24_{-0.54}^{+0.47}$ & $6.73_{-0.28}^{+0.24}$ & $50.33_{-0.02}^{+0.02}$ & $49.06\pm0.01$ & $2.67_{-0.07}^{+0.05}$ & $2.54\pm0.05$ & $3.48_{-0.03}^{+0.04}$ & $3.48\pm0.03$\\
180329B & $68.27_{-0.38}^{+0.49}$ & $7.93_{-0.16}^{+0.20}$ & $48.87_{-0.04}^{+0.02}$ & $47.41\pm0.01$ & $3.58_{-0.25}^{+0.16}$ & $3.04\pm0.17$ & $3.77_{-0.06}^{+0.11}$ & $3.91\pm0.07$\\
180404A & $55.99_{-0.43}^{+0.50}$ & $4.07_{-0.22}^{+0.26}$ & $47.85_{-0.06}^{+0.07}$ & $46.26\pm0.02$ & $4.71_{-0.53}^{+0.68}$ & $4.10\pm0.42$ & $3.98_{-0.28}^{+0.19}$ & $3.89\pm0.22$\\
\hline
\hline
\end{tabular}
\label{table3}
}
\end{table*}

\section{Conclusions and Discussion}
\label{sec4}
In this work we have revisited the scenario that a GRB X-ray afterglow is powered by a continuous energy injection of a newborn magnetar. Generally we would expect to test the internal magnetic dissipation model via spectral evolution from observations. For the canonical parameters used in Figure \ref{fig1}, at the beginning the observed frequency of {\it Swift}-XRT satisfies $\nu_c<\nu_a<\nu_{XRT}<\nu_m$, and then turns into $\nu_a<(\nu_m,\,\nu_{XRT}) <\nu_c$ quickly. As indicated by Eq. (14-16) of our previous work \citep{XiaoD2017}, at the beginning the X-ray spectrum should be $L_\nu \propto \nu^{-0.5}$ and becomes $L_\nu \propto \nu^{-(p-1)/2}$ later. To identify this spectral evolution from observations, we need the spectrum measurement at very early times (at the beginning of the plateau phase), otherwise the spectral index has become $-(p-1)/2$. For the power-law component that has an external origin, the X-ray spectral index should be $-(p-1)/2$ or $-p/2$, depending on whether the electrons accelerated by external shocks are in slow or fast cooling regime. For most of the cases in XRT archival data, the X-ray spectral indexes are around $ -(p-1)/2$ during the plateau phase and it is not easy to test the model. However, we still have some special cases like GRB 061121, whose X-ray photon index shows a clear transition from $1.46\pm0.03$ at T0+173s (very close to the beginning of plateau phase) to $1.83\pm0.06$ at T0+20872s \citep{Evans2009}. This spectral evolution from $\sim -0.5$ to $ -(p-1)/2$ strongly favors the model presented in this work. Moreover, a general prediction of this model is that there will be gamma-ray emission at the same time of X-ray plateau phase. However, since the spectral index is $ -(p-1)/2$, the gamma-ray flux is much lower than simultaneous X-ray flux (typically $\leq10^{-9}\,\rm erg\,cm^{-2}\,s^{-1}$, as we can see from Figure 4). Such a low gamma-ray flux is well below the detection threshold of {\it Swift}-BAT and {\it Fermi}-GBM so that it is not easy to be observed.

For those X-ray plateaus that have ``internal" origins, we start from the radiation process induced by the magnetic energy dissipation within the magnetar wind to calculate the X-ray radiation efficiency. This approach is much more realistic and reasonable than the commonly assumed constant efficiency. We have found that the X-ray radiation efficiency depends strongly on the injected luminosity. This relation has an important impact on the temporal decay index after the plateau phase, namely, making $\beta$ deviate from $-2$. This implies that the requirement of braking index $n<3$ for $\beta<-2$ type light curve in previous studies \citep[e.g.,][]{ZhangB2001, Lasky2016, LvHJ2018} is now largely relieved.

Moreover, this relation has a straightforward implication that the braking index deduced from afterglow light curve fitting should be reconsidered with care. For illustration we have adopted the exactly same sample as in \citet{LvHJ2018} and redone the fitting procedure. The braking indexes are clearly larger and the number of cases with $n<3$ is much less than that of their work. This should be closer to reality for newborn magnetars, since $k$ is likely to remain unchanged in such a short time ($<10^6$ s). The effects causing an evolving $k$ (e.g., the evolution of magnetic field strength or angle between rotation axis and dipole field axis) are expected to happen on a much longer timescale \citep[e.g.,][]{ChenWC2006}.

The X-ray radiation efficiency depends strongly on the saturation Lorentz factor, which is equivalent to initial magnetization \citep{Beniamini2017b}. For a typical value of $\sigma_0=100$ usually adopted in GRB study, $\eta_X$ is of order $10^{-2}$ and much smaller than the value $0.1$ usually assumed in previous studies. This means that previous studies may underestimate the initial magnetic dipole luminosity $L_0$. This will in turn limit the initial spin period and magnetic field strength. In this sense, constraining $\sigma_0$ from X-ray afterglow plateau may be possible. All of these need to be reevaluated carefully in future work.

\acknowledgements
We are grateful to Bin-Bin Zhang for providing the MCurveFit package, which is essential in generating the results of this paper. This work was supported by the National Key Research and Development Program of China (Grant NO. 2017YFA0402600) and the National Natural Science Foundation of China (Grant NO. 11573014, 11833003, and 11851305). D.X. was also supported by the Natural Science Foundation for the Youth of Jiangsu Province (Grant NO. BK20180324).

\bibliographystyle{aasjournal}
\bibliography{bb}

\end{document}